# Chaotic principle: some applications to developed turbulence[*]

*Giovanni Gallavotti*

Fisica, Università di Roma *La Sapienza*
P.le Moro 2, 00185, Roma, Italia.

**Abstract:** *Some models for developed turbulence are considered; they are shown to obey a large fluctuations theorem, and one among them also obeys a response reciprocity relation of Onsager's type. This illustrates and extends ideas and techniques developed in earlier works mainly for non equilibrium problems in statistical mechanics.*

§1. **Introduction. Earlier results.**

The question: "what are the measures describing turbulence?" has been repeatedly raised in a clear form by Ruelle, *e.g.* see [R1], p. 6÷8, who proposed on many occasions that the probability distributions describing turbulence should share some selected properties among the many ones enjoyed by the SRB distributions for axiom A attractors: and in fact it has become customary to call SRB distributions the probability distributions on phase space that describe the statistics of turbulent (chaotic) motions, [ER].

It does not seem that Ruelle ever wrote explicitly how the above prescription could be actually implemented for testing: but his idea was made quite clear through his writings and seminars. Nevertheless the generality and the breadth of the proposal were never really picked up with the purpose of obtaining concrete predictions.

Recently the following interpretation of Ruelle's prescription for describing turbulence and, more generally, motions having an empirically chaotic nature has been proposed, [GC1],[GC2]:

*Chaotic hypothesis: A reversible many particle system in a stationary state can be regarded as a transitive Anosov system for the purpose of computing macroscopic properties.*

See [AA],[S],[R2],[Bo] for the notion and properties of Anosov systems. In the very common case of *non reversible* systems one has to replace the transitive Anosov propery with the more general Axiom A system property, [R2].

The time evolution is described by a flow $V_t$ generated by differential equations on a *continuous phase space* $\mathcal{F}$.

However we shall always regard the time evolution, or the *dynamics*, as a map $S$ acting on the "phase space" $\mathcal{C}$ of the *observed events*, also called *timing events*, which could be, for instance, the occurrence of a microscopic binary "collision" in a particle system or the event in which a prefixed component of the velocity field assumes a certain value, or a maximum value, in fluid motions (*e.g.* a typical example is Lorenz' choice for timing the observations via the maxima of the variable he was calling $Z$, [L]). If $x \in \mathcal{C}$ then $V_t x$ evolves in $\mathcal{F}$ until the next timing event: $Sx = V_{t(x)} x \in \mathcal{C}$, if $t(x)$ is the time elapsing between the timing event $x$ and the next.

---

The above chaotic hypothesis implies that the macroscopic time averages of observables are described by a probability distribution $\mu$ on the "phase space" $\mathcal{C}$ of the *observed events*, also called *timing events*, (which could be, for instance, the occurrence of a microscopic binary "collision").

The existence of the distribution $\mu$ is assumed *a priori* in general, as stated by the following (extension, see [GC2]) of the zeroth law, [UF], giving a global property of the motions generated by initial data chosen randomly with distribution $\mu_0$ proportional to the volume measure on $\mathcal{C}$:

*Extended zero-th law: A dynamical system $(\mathcal{C}, S)$ describing a many particle system (or a continuum such as a fluid) describes motions that admit a statistics $\mu$ in the sense that, given any (smooth) macroscopic observable $F$ defined on the points $x$ of the phase space $\mathcal{C}$, the time average of $F$ exists for all $\mu_0$-randomly-chosen initial data $x$ and is given by:*

$$\lim_{T \to \infty} \frac{1}{T} \sum_{k=0}^{T-1} F(S^j x) = \int_{\mathcal{C}} \mu(dx') F(x') \tag{1.1}$$

*where $\mu$ is a $S$-invariant probability distribution on $\mathcal{C}$.*

If one assumes the chaotic hypothesis, then it *follows* that the zeroth law holds, [S,Bo,R2]; however it is convenient to regard the two statements as distinct because the hypothesis we make is "*only*" that one can suppose that the system is Anosov for "practical purposes": this leaves the possibility that it is not strictly speaking such and some corrections may be needed on the predictions obtained by using the hypothesis. The corrections are supposed to become neglegible in the thermodynamic limit (in statistical mechanics) or in the large Reynolds numbers regimes (in fluids): but the hypothesis, in our intentions, has a meaning similar to that of the ergodic hypothesis. The latter too is supposed to be very reliable (*under general circumstances* as there are well known exceptions, *e.g.* the free gas in statistical mechanics or some very special Navier Stokes flows, [M]) in the mentioned limiting situations: but it is very often *as reliable* in systems with few degrees of freedom.

Still the above principle may look at first sight quite abstract and, in the end, useless. Our idea in [GC1],[GC2] was that *instead* it could eventually reveal itself comparably powerful to the ergodic hypothesis in equilibrium non dissipative systems: we began investigating whether this statement could be verified in some concrete cases of non equilibrium systems governed by a *reversible* dissipation mechanism and the following results can be obtained.

*(1) The fluctuation theorem* (in reversible dissipative systems): The energy dissipation or more precisely the *entropy creation* per timing event $\sigma(x)$ in the configuration $x$, identified with the logarithm of the inverse of the *phase space volume contraction* $e^{-\sigma(x)}$, has a (future) time average $\langle \sigma \rangle_+$, positive by assumption, but it is a fluctuating variable in the models considered. Therefore one can consider the average $\langle \sigma \rangle_\tau(x)$ of $\sigma(x)$ evaluated on a time interval of $\tau$ units during which the system evolves between $S^{-\tau/2} x$ and $S^{\tau/2} x$:

$$\sigma_\tau(x) = \frac{1}{\tau} \sum_{k=-\tau/2}^{\tau/2-1} \sigma(S^k x) \tag{1.2}$$

Note that the timing events occur at varying time intervals: the time between the event $x \in \mathcal{C}$ and $Sx$, denoted above $t(x)$, has an average value $t_0 = \langle t(\cdot) \rangle$ and the entropy generation rate per unit physical (*i.e.* continuous) time is $t_0^{-1} \sigma(x)$.

Reversibility is not contradictory with the presence of dissipation: see [PH], [ECM1], [CG2], and the models considered in the present work.

We can study the probability distribution of this random variable, with respect to the SRB distribution of the motion, *i.e.* in the stationary state. If we introduce, for convenience, a dimensionless fluctuation variable $p$ by setting:

$$\sigma_\tau(x) = \langle \sigma \rangle_+ \, p \tag{1.3}$$

where $p$ will have (therefore) SRB average 1.

Then the main result of [ECM2],[GC1],[GC2] is a general property of the *probability distribution of* $p$ denoted $\pi_\tau(p) \equiv e^{-\tau \zeta_\tau(p)}$, with respect to the stationary statitics $\mu$ (*i.e.* with respect to the SRB distribution):

$$\lim_{\tau \to \infty} \frac{1}{p\,\tau\langle\sigma\rangle_+} \log \frac{\pi_\tau(p)}{\pi_\tau(-p)} = 1 \qquad \text{for all } p \tag{1.4}$$

The above property has *no free parameters* and it can be tested in some numerical experiments: in fact it was discovered in the experiment on particle systems in [ECM2] inspired by the literature on the SRB distributions; in [ECM2] the connection with the Ruelle's proposals was also, sketchily, worked out. The detailed discussion of the connection between the chaotic hypothesis and the *fluctuation theorem* (1.4) is presented in [GC1],[GC2], with some mathematical details worked out in [G1],[G2].

The above result applies as well to certain fluid mechanics flows proposed first in [ZJ] and mentioned in [GC1],[GC2] and discussed again below, for completeness. The consideration of this kind of models goes back to older ideas, see [PH] for a review, which inspired also [ECM2], [GC1], [GC2].

*(2) Onsager reciprocity:* suppose that the system is subject to external *thermodynamic forces*, *i.e.* its equations of motion depend on parameters $a, b, c...$ such that the system is hamiltonian when the parameters values are 0 and becomes dissipative but *still reversible*, when they are different from 0. The *response current*, $J$, to variations of the parameters is defined together with its SRB average $j$ as:

$$J_a(x) = \frac{1}{t_0} \frac{\partial \sigma(x)}{\partial a}, \qquad j_a = \langle J_a \rangle_{SRB} \tag{1.5}$$

where $t_0$ is the average time between timing events.

Then, if the total energy of the system is kept constant, the reciprocity relation is:

$$\partial_b j_a \big|_{a,b=0} \stackrel{def}{=} L_{ab} = L_{ba} \stackrel{def}{=} \partial_a j_b \big|_{a,b=0} \tag{1.6}$$

and it can be obtained, from the above chaotic hypothesis, in the case of various statistical mechanics models, [CG]. For a general discussion of Onsager's reciprocity see [DGM], [ELS].

Result (1) is a *large deviation theorem* and (therefore) it can be tested only in systems with few degrees of freedom (because $\zeta_\tau$ is proportional to the number of degrees of freedom). It is, however, a prediction.

Result (2) is, instead, an *a priori* test of the hypothesis: in fact reciprocity is independently known to hold (for macroscopic systems). It is, therefore, not a prediction but a check if consistency with the body of the results that are considered independently established, see [DGM].

In this paper the above large deviation theorem (already derived as the main result in [GC1], [GC2]) is reproduced in the present context, for completeness as well as to introduce the basic ideas and notations associated with the chaotic hypothesis. Then we investigate the extension of the result (2) to another model related to the incompressible Navier Stokes equation in the Kolmogorov Obuchov scaling regime.

## §2. A model for developed turbulence.

We consider the Navier Stokes (NS) equations in a box $[-\frac{L}{2}, \frac{L}{2}]^3$, with periodic boundary conditions and for an incompressible fluid. If the velocity field is written in a Fourier series as:

$$\underline{u}(\underline{x}) = \sum_{\underline{k} \neq \underline{0}} e^{i\underline{k}\cdot\underline{x}} \underline{\gamma}_{\underline{k}} \tag{2.1}$$

with $\underline{\gamma}_{\underline{k}}$ complex vectors with $\underline{\gamma}_{\underline{k}} = \overline{\underline{\gamma}}_{-\underline{k}}$ (reality of the velocity field) and $\gamma_{\underline{k}} \perp \underline{k}$ (incompressibility) then (it is well known and easy to check that) the NS equations become:

$$\dot{\underline{\gamma}}_{\underline{k}} = -i \sum_{\underline{k}_1 + \underline{k}_2 = \underline{k}} (\underline{\gamma}_{\underline{k}_1} \cdot \underline{k}_2) \Pi_{\underline{k}} \underline{\gamma}_{\underline{k}_1} + R\underline{g}_{\underline{k}} - \nu \underline{k}^2 \underline{\gamma}_{\underline{k}} \tag{2.2}$$

where $\Pi_{\underline{k}}$ is the orthogonal projection over the plane orthogonal to $\underline{k}$; $\nu$ is the kinematic viscosity and $R\underline{g}_{\underline{k}}$ is the forcing (of course orthogonal to $\underline{k}$) which will be taken to be non zero only for a few components with small $\underline{k}$. Since $\underline{k} = \frac{2\pi}{L} \underline{n}$ with $\underline{n}$ integer this means that the force acts only on the high length scale components.

For simplicity we may think that the forcing has only two non vanishing components $R\underline{g}_{\underline{k}_1^0}$, $R\underline{g}_{\underline{k}_2^0}$, corresponding to two linearly independent wave numbers $\underline{k}_1^0, \underline{k}_2^0$. The simpler case of only one non zero component can be trivial (*e.g.* in 2 dimensional NS when the forcing acts on the smallest $\underline{k}$, $|\underline{k}| = k_0$, [M]), and is therefore discarded here in favor of the next to the simplest. *We shall keep $\underline{g}$ fixed throughout the analysis* and we set $g = \max|\underline{g}_{\underline{k}}|$.

The number $R$ therefore determines the forcing strength and will be identified with the Reynolds number (we also keep the container size $L$ and the viscosity $\nu$ fixed). We take $\underline{\gamma}_{\underline{0}} \equiv \underline{0}$ since it is the conserved center of mass velocity.

In order to obtain equations in the framework of this paper from the phenomenological theory of Kolmogorov–Obuchov, [LL], we shall assume that the above equations can be replaced by the following simpler ones:

$$\begin{aligned}\dot{\underline{\gamma}}_{\underline{k}} &= -i \sum_{\underline{k}_1 + \underline{k}_2 = \underline{k}} (\underline{\gamma}_{\underline{k}_1} \cdot \underline{k}_2) \Pi_{\underline{k}} \gamma_{\underline{k}_1} + \underline{g}_{\underline{k}} & |\underline{k}| < k_R \\ \dot{\underline{\gamma}}_{\underline{k}} &= -\alpha \underline{\gamma}_{\underline{k}} - i \sum_{\underline{k}_1 + \underline{k}_2 = \underline{k}} (\underline{\gamma}_{\underline{k}_1} \cdot \underline{k}_2) \Pi_{\underline{k}} \gamma_{\underline{k}_1} & k_R \leq |\underline{k}| < k_R + (RLg)^{1/2}\nu^{-1}\end{aligned} \tag{2.3}$$

where, if $k_0 = \frac{2\pi}{L}$, the wave vector $k_R$ is the Kolmogorov momentum scale $k_R = k_0 R^{3/4}$, ([LL] p. 122, (32.6)). We also set $N_R$ to be the number of wave vectors pairs ("modes") $\underline{k}, -\underline{k}$ such that $k_R \leq |\underline{k}| \leq k_R + (RLg)^{1/2}\nu^{-1}$, then $N_R \approx (\frac{k_R}{k_0})^2 \frac{(RLg)^{1/2}\nu^{-1}}{k_0} \approx R^2$.

Then the total number of pairs $\underline{k}, -\underline{k}$ of modes considered is $N \approx R^{9/4}$ among which $N_R \approx R^2$ are *viscous modes*, while the others are *inertial modes*. The number of independent components of the field $\underline{\gamma}_{\underline{k}}$ is $4N$ (recall also that, taking into account the reality and incompressibility conditions, forces the $\underline{\gamma}_{\underline{k}}, \underline{\gamma}_{-\underline{k}}$ to have only two linearly independent *complex* components). We call (2.3) the *KO model*.

This means that the equations for the amplitudes $\underline{\gamma}_{\underline{k}}$ corresponding to $\underline{k}$'s in the *inertial range*, $k_0 \leq |\underline{k}| \leq k_R$, are "governed" by the reversible Euler equations. In the *viscous range*, $|\underline{k}| > k_R$ the dissipation phenomena will be idealized by saying that the equations are simply such that only the modes $\underline{k}$ with $k_R < |\underline{k}| < k_R + (RLg)^{1/2}\nu^{-1}$ have a non zero amplitude and evolve in such a way as to keep the total energy constant. This means that the parameter $\alpha$ plays the role of an effective thermostat (or viscosity), which has to be chosen so that the total energy is constant, *i.e.* so that $\frac{d}{dt}\sum_{\underline{k}}|\underline{\gamma}_{\underline{k}}|^2=0$:

$$\alpha(x)=\frac{\sum_{\underline{k}}\underline{f}_{\underline{k}}\cdot\underline{\gamma}_{-\underline{k}}}{\sum_{|\underline{k}|>k_R}|\underline{\gamma}_{\underline{k}}|^2} \tag{2.4}$$

The phase space contraction rate is therefore the divergence of the r.h.s. of (2.3), *i.e.* $4N_R\alpha(x)$. Hence the entropy generation per per timing event is $\sigma(x)$ such that:

$$\frac{1}{t_0}\sigma(x)=4N_R\alpha(x)=\frac{\sum_{\underline{k}}\underline{f}_{\underline{k}}\cdot\underline{\gamma}_{-\underline{k}}}{(4N_R)^{-1}\sum_{|\underline{k}|>k_R}|\underline{\gamma}_{\underline{k}}|^2}\stackrel{def}{=}\frac{\varepsilon(x)}{kT(x)/2} \tag{2.5}$$

where $t_0$ is the average time between timing events and $\varepsilon(x)$ and $kT(x)/2$ are simply *defined* respectively by the numerator and denominator of the above fraction defining $\alpha(x)$.

In (2.5) and in the following we neglect, *for simplicity*, the variability of the time $t(x)$ elapsing between timing events. If one wished to avoid this approximation the r.h.s. of (2.5) should be replaced by $t_0^{-1}$ times its integral over the continuous time trajectory over $[0,t(x)]$ of $4N_R\alpha(V_t x)$ (using the notations fixed in §1).

The Kolmogorov length $k_R^{-1}$ is introduced here phenomenologically and we do not attempt at a fundamental derivation of (2.3), (2.4). Therefore (2.3) has to be regarded as a phenomenological equation. A class of similar models was introduced in [ZJ].

Note that $\sigma$ is proportional to the work $\varepsilon(x)$, per unit time and per viscous degree of freedom, performed on the system and dissipated into heat in order to keep the total energy constant: the proportionality constant is $2/kT(x)$ with $\frac{1}{2}kT(x)\equiv\frac{1}{4N_R}\sum_{|\underline{k}|<k_R}|\underline{\gamma}_{\underline{k}}|^2$ (which, however, is *not* a constant on the motions described by (2.3),(2.4) because of the imposed constraint that $\sum_{\underline{k}}|\underline{\gamma}_{\underline{k}}|^2$ is constant, rather than $\sum_{|\underline{k}|>k_R}|\underline{\gamma}_{\underline{k}}|^2$).

Hence $\langle\sigma\rangle_+$ can be thought of as the average amount of *energy dissipation* per unit time by the flow divided by the *kinetic energy* per mode contained in the *viscous modes*. The first quantity plays a major role in Kolmogorov's theory, see [LL] p.119, and its average is usually called $\varepsilon$, (see [LL], (31.1)). Since the kinetic energy contained in the viscous modes can be considered as a kind of "temperature" we see that $\langle\sigma\rangle_+$, is proportional to an entropy "production rate". More appropriately we can say that, for $R$ large, $\langle\sigma\rangle_+$ is proportional to the "energy dissipation rate" over a kinetic quantity equal to the average kinetic energy contained in the viscous modes *provided*, for large $R$, the two quantities can be regarded as independent random variables (so that the average of this ratio equals the ratio of their averages).

Note that for the above model (2.3) the time reversal map $i:\{\underline{\gamma}_{\underline{k}}\}\to\{-\underline{\gamma}_{\underline{k}}\}$ has the property that $iV_t=V_{-t}i$ if $V_t$ is the flow generated by the (2.3): see §1 for the notation.

As discussed in §1, we shall call $\mathcal{C}$ the space of the velocity fields $\underline{\gamma}$ with a fixed value $U$ for the total kinetic energy and such that a fixed velocity component (*e.g.* an arbitarily selected viscous component $\gamma_{\underline{k}_1,1}$ in the $x$–direction) has a given value (or a local maximum, to follow [L]). We call the space $\mathcal{C}$ the space of the *timing events* as we shall imagine to record the velocity field every time one event $x$ in $\mathcal{C}$ occurs. Then the time evolution flow induces a map $S$ on $\mathcal{C}$ mapping one element $x\in\mathcal{C}$ into the event of the same type which occurs after it. If $t(x)$ is the time between the timing event $x$ and the next following it it is $Sx=V_{t(x)}x$.

The time reversal symmetry for the continuum flow becomes the symmetry $Si=iS^{-1}$ on the timing map $S$ that we shall call the evolution map.

A similar model is the one considered originally by [ZJ]: in this model the equations of motion are given by the (2.3) but with the thermostatting forces $-\alpha\underline{\gamma}_{\underline{k}}$ present also in the first equation in (2.3). The value of $\alpha$ is still determined by imposing the constancy of the total kinetic energy. Therefore the (2.4) is modified by replacing the sum in the denominator by an unrestricted sum. This means that in this model $T(x)$ is rigorously constant and therefore there is a rigorous proportionality between the phase space contraction rate and the energy dissipation rate.

### §3. Chaotic hypothesis and vanishing Lyapunov exponents.

In the above models the dimension of the phase space $\mathcal{C}$ on which the time evolution is represented by a map between timing events is two units less than the dimension of the space of the velocity fields (where the motion is described by differential equations like (2.3)).

This reduction is useful not only because it eliminates a degree of freedom which is "trivial" (by fixing the energy which is, in any event, constant) but also because it eliminates the degree of freedom corresponding to the direction of the flow which is responsible for the existence of a 0 Lyapunov exponent.

Obviously had we not assumed the point of view of regarding the time evolution as a map on $\mathcal{C}$, then this would have caused us the problem that the chaotic hypothesis would have been in conflict with a basic property of Anosov systems: in fact the Lyapunov exponents for such systems are separated from 0 by a gap $\lambda > 0$.

But one may well have doubts that considering the evolution on a space of timing events is sufficient to avoid that the chaotic hypothesis comes into a manifest contradiction because of the possible existence of *other* vanishing Lyapunov exponents.

For instance in some examples, *e.g.* model 2 in [GC2], the dimension is reduced by 4 units because the equations of motion preserve a component of the center of mass momentum, *provided it has 0 value*, and also the corresponding center of mass position coordinate. If the total momentum is not initially 0 it relaxes to the value 0 with a negative Lyapunov exponent, while the center of mass motion relaxes to a rectilinear motion with zero exponent.

Therefore fixing the values of the total momentum component and the center of mass coordinate further reduces the dimension by 2 and eliminates two more "trivial coordinates".

In this case the conservation of the horizontal momentum is due to the special boundary conditions used: it was pointed out in [GC2] that changing the boundary conditions may turn the horizontal momentum and the center of mass position into a non exactly conserved quantity, no matter which is the value given to it initially. And in such circumstances we can expect that while the momentum evolves towards its stationary value (0) at exponential rate (*i.e.* "with a non zero Lyapunov exponent") the slow almost linear motion of the center of mass generates a *vanishing* Lyapunov exponent, at least in the limit of large systems).

But one should expect that when the system becomes large (*i.e.* in the *thermodynamic limit*) the boundary conditions should have neglegible influence on the macroscopic properties of the system. Therefore the macroscopic dynamics may be insensitive to the above variability and, in consequence, adding *as a constraint* that the horizontal momentum vanishes exactly, hence the center of mass position is also on rectilinear motion exactly, *should not* affect the macroscopic behaviour. At least if the forces needed to impose the constraint are provided by a *minimal constraint* principle (see below).

One should note that the above replacement of an almost conserved quantity by an exactly conserved one has in fact the consequence of eliminating a trivial *pair* of Lyapunov exponents: one is 0 and one is negative and describes the relaxation to equilibrium of a macroscopically interesting quantity (the total horizontal momentum, *i.e.* the horizontal current, in the mentioned example).

In [GC2] it was proposed, see §8, that the above mechanism might be quite general. In the sense that a non equilibrium system may have *many*, perhaps *very many* 0 Lyapunov exponents (evidence in this direction can be found in the basic paper [LPR]): this does therefore (apparently) violate the chaoticity hypothesis of §1 strongly.

In such systems the chaotic hypothesis can still be assumed, it was suggested in [GC2] §8, if the dynamical variables responsible for the existence of the 0 Lyapunov exponents can be identified togheter with the conjugate variables (which would have negative Lyapunov exponents, and which are expected to exist quite generally at least if one accepts as generally valid the pairing rule discovered in [ECM1] and discussed in [EM], [ES], [GC2]), and then one modifies the equations of motion so that the identified variables are exactly constant (with the one with negative Lyapunov exponent fixed at its equilibrium value).

The proposal in [GC2] was that the minimal forces necessary to impose the constraint would not affect the macroscopic behaviour of the motions: here by minimal forces we proposed to intend the forces prescribed by *Gauss principle of least constraint*, (see [LA], vol. $II_2$, p. 470, and appendix A1 below). The idea being that the dynamics enforces the constraint whether it is present or not, *provided the constrained variables are assigned the appropriate stationary average value*.

The chaotic hypothesis can then be used to describe the evolution of the remaining coordinates, if one has taken into account all the macroscopic constraints so that no 0 Lyapunov exponents are present any more.

The question is whether one can identify concretely all the dynamical variables that can be fixed to have a well defined value without affecting the macroscopic behaviour of the system.

A guide to finding such variables should be the macroscopic equations that the system is supposed to obey. To give an example of what I have in mind consider a dense gas which can be regarded as an incompressible inviscid fluid. In this case we can reasonably expect, [EMY,KV,LOY], that the particles motion proceeds in such a way that the macroscopic average velocity field $\underline{u}(\underline{x})$ evolves according to the Euler equations, while the displacement field, relative to the initial configuration, changes accordingly.

Therefore *given the solution of the Euler equations* corresponding to the initial macroscopic state of the gas: $t \to \underline{u}(\underline{x},t)$, one can impose that the average velocity locally coincides with $\underline{u}(\underline{x},t)$. This means that we can consider a lattice of cubes $\Delta_\gamma$ with microscopically large and macroscopically small side, see [KV], and call $\underline{x}_\gamma$ the center of the cubes; then we can impose the constraints:

$$\sum_j \chi_\gamma(\underline{q}_j)\frac{\underline{p}_j}{m} = \underline{u}(\underline{x}_\gamma,t), \qquad \frac{1}{2}\sum_j \chi_\gamma(\underline{q}_j)(\frac{\underline{p}_j}{m} - \underline{u})^2 = N_\gamma kT \qquad (3.1)$$

where $T$ is the temperature and $\chi_\gamma$ is the characteristic function of $\Delta_\gamma$, via the imposition of auxiliary forces derived from Gauss minimal constraint principle (see [LA]), *i.e.* by writing the equations:

$$\underline{\dot q}_j = \frac{1}{m}\underline{p}_i, \qquad \underline{\dot p}_j = \underline{f}_j + \sum_\gamma \left(\underline{\alpha}_\gamma + \beta_\gamma(\frac{\underline{p}_j}{m} - \underline{u})\right)\chi_\gamma(\underline{q}_j) \qquad (3.2)$$

where $\underline{\alpha}_\gamma$ are determined by imposing that the constraints are exactly verified. This is:

$$\underline{\alpha}_\gamma = \underline{\dot u}(\underline{x}_\gamma, t) \qquad (3.3)$$

and $\beta_\gamma$ is determined likewise.

One expects that in this way a very large number of coordinates which evolve with a negative Lyapunov exponent is eliminated; and that the corresponding position coordinates, *i.e.* the coordinates of the displacements with respect to the initial positions, change with a 0 Lyapunov exponent.

Some care has to be exercized here: the remaining coordinates will be supposed to vary with Lyapunov exponents separated by a gap $\lambda$ from the value 0, independently of the system size. This means that the evolution of the very large number of remaining coordinates takes place on a very fast time scale $\lambda^{-1}$. We can expect the latter to be the same time scale on which the local averages reach the "equilibrium" value given by $\underline{u}(\underline{x},t)$ (this is a consequence of the *pairing rule*, [ECM1],[GC2], if accepted). On the other hand the displacement variables should evolve with a 0 Lyapunov exponent: but the latter is 0 only *if compared to* $\lambda$. It may well be *non zero and very small* so that its inverse is a macroscopic time, as it is shown in many numerical experiments (that show positive Lyapunov exponents of the "lagrangian motions").

This would mean that the chaotic hypothesis is valid for finite systems, in the space of the timing events, and that it becomes invalid strictly speaking only in the thermodynamic limit, when however it is not valid for trivial reasons and assuming it simply corresponds to think that some approximate conservation laws have become exact macroscopic evolution laws thus implying the vanishing of some Lyapunov exponents. The phenomenon can be avoided by imposing, already in finite systems, the approximate conservation laws *as exact laws* and eliminating the relative coordinates (thus restoring a uniform gap in the Lyapunov spectrum).

In our fluid model the situation is likely to be similar: and one has to interpret accordingly the chaotic hypothesis: namely one has to think that there are no 0 Lyapunov exponents *or* that if they are present they can be eliminated by adding extra forces that fix the value of some observables which would relax slowly to equilibrium (thus generating vanishing Lyapunov exponents) without affecting the behaviour of the system (except of course for what concerns the long time correlations of the very same observables evolving with vanishing Lypapunov exponents).

As an example of the above discussion one can argue that the model (2.3) and its modification in which the thermostatting forces act on all modes (considered at the end of §2, [ZJ]) can be considered without obvious contradictions to verify the chaotic hypothesis for finite $R$, and uniformly in $R$ at least as far as the properties of the observables that relax quickly to equilibrium are concerned.

### §4. The SRB distribution.

The application of the chaotic hypothsis, as we proposed in [GC1],[GC2], is very similar to the applications of the ensembles method in equilibrium statistical mechanics. In that case one does not really need to evaluate all the monstrous integrals over phase space to deduce remarkable macroscopic properties: a great example is provided by Boltzmann's derivation of the heat theorem, [B], [G3], *i.e.* of macroscopic thermodynamics.

The chaotic hypothesis can perhaps be used in a similar way because it provides us (see below) with an apparently impossibly complicated expression for the SRB distribution: it is nevertheless an expression on which, as we have shown in [GC1],[GC2], one can work quite in detail and from which the concretely testable propeties discussed in §1 have been derived.

The key point is that the hypothesis implies the possibility of defining a *natural coarse graining* of the phase space which is also *mathematically precise* (a problem that has been previously repeatedly debated without ever reaching clear conclusions, in my understanding at least, [G4]).

In fact in an transitive Anosov (or axiom A) system one can define a partition $\mathcal{E}$ of phase space $\mathcal{C}$ into $\mathcal{N}$ sets $E_1, E_2, \ldots, E_\mathcal{N}$ which are "regularly" shaped (*i.e.* with non empty interior, with boundary which has zero volume and which is smooth at least in the sense of Hölder continity).

The sets in $\mathcal{E}$ have the (highly non trivial) property that if one defines a *compatibility matrix* $C_{ij}$ by setting $C_{ij} = 1$ if the interior of $SE_i$ intersects the interior of $E_j$ and 0 otherwise, then any sequence $\underline{j} = (j_k)_{k=-\infty,\infty}$ such that $C_{j_k,j_{k+1}} = 1$ for all $k$'s (*compatible sequence*) is the *history on $\mathcal{E}$ of one and only one point $x \in \mathcal{C}$*, in the sense that $S^k x \in E_{j_k}$ for all $k$'s.

Viceversa each point $x \in \mathcal{C}$ is generated by one compatible sequence, naturally called the *history* of $x$ on $\mathcal{E}$, in the sense that it is the only point in the intersection $\cap_k S^{-k} E_{j_k}$. There may be more sequences determining the same point but this happens for a zero volume set of points (this is a quite trivial ambiguity as it is similar (and in fact very closely related to) the well known ambiguity that one has in representing the reals in basis 10 by digits that end in a sequence of 9's or of 0's). Furthermore the matrix $C$ admits a power $C^q$ with all the entries positive (this is a consequence of the transitivity).

The above property allows us to think of the compatible sequences as a representation of our phase space and to regard the volume measure as a measure on the space of the compatible sequences. The SRB distribution $\mu$ also can be regarded as a probability distribution on the space of the compatible sequences. The latter, in turn, can be conveniently regarded as the space of the spin configurations of a one dimensional spin model (the *spin at the site k* being $j_k$).

The distribution $\mu$ has a remarkable representation when considered as a distribution on the space of the compatible sequences,

To describe the representation we need some further notations. Given a symbol $j$ we can find a semiinfinite compatible sequence $\underline{j}_+ = (j, j', j'', \ldots)$ whose entries depend only on the value of $j$; likewise we can find a seminfinite compatible sequence $\underline{j}_- = (\ldots, j'', j', j)$ whose entries depend only on $j$. This property, consequence of the above mentioned transitivity, can be used to locate conveniently a point whose symbolic representation contains a string $\underline{j}_{a,b} = j_a, j_{a+1}, \ldots, j_b$: one simply continues the string $\underline{j}_{a,b}$ into the infinite string: $\underline{j} = (\underline{j}_-(j_a), \underline{j}_{a,b}, \underline{j}_+(j_b))$ obtained by merging sequentially the three strings $\underline{j}_-(j_a)$, $\underline{j}_{a,b}$ and $\underline{j}_+(j_b)$.

Let $E = E_{j_{-M}, \ldots, j_M} = \cap_{k=-M}^{M} S^{-k} E_{j_k}$: this is the set of points $x$ such that $S^k x \in E_{j_k}$ for $k = j_{-M}, \ldots, j_M$. We fix a point $x_E \equiv x_{j_{-M}, \ldots, j_M} \in E_{j_{-M}, \ldots, j_M}$ to be the point whose history is obtained by continuing arbitrarily the sequence $j_{-M}, \ldots, j_M$ "to the right and to the left" into a biinfinite compatible sequence $\underline{j}$ as described in the previous paragraph. The point $x_E$ will be called the *center* of $E$.

The expansivity of an Anosov map implies that the sets $E = E_{j_{-M}, \ldots, j_M}$ are very small (their diameter tends to 0 as $e^{-\lambda M}$ if $\lambda$ is, essentially, the gap in the Lyapunov spectrum) and we can define a distribution $\mu_M$ by assigning a weight to each of such sets.

The weight that we assign to them is related to the expansion coefficient $\overline{\Lambda}_{u,M}(x_E)$ of the map $S^M$ as a map between $S^{-M/2} x_E$ and $S^{M/2} x_E$.

The expansion (contraction) coefficient $\Lambda_u(x)$ (respectively $\Lambda_s(x)$) of $S$, at $x$, is the jacobian determinant (evaluated at $x$) of the transformation $S$ as a map of the unstable (stable) manifold $W^u(x)$ ($W^s(x)$) into itself. For a discussion of the notion of stable and unstable manifolds see [R1], [ER]. Therefore the expansion (contraction) coefficient $\overline{\Lambda}_{u,M}(x)$ (respectively $\overline{\Lambda}_{s,M}(x)$) are given by:

$$\overline{\Lambda}_{u,M}(x) = \prod_{j=-M/2}^{M/2-1} \Lambda_u(S^j x) \stackrel{def}{=} e^{M \overline{\lambda}_{uM}(x)}, \qquad \overline{\Lambda}_{s,M}(x) = \prod_{j=-M/2}^{M/2-1} \Lambda_s(S^j x) \stackrel{def}{=} e^{-M \overline{\lambda}_{sM}(x)} \quad (4.1)$$

and the distribution $\mu_M$ is defined by giving to each $E_j \in \mathcal{E}_M$, with center $x_{E_j} \equiv x_j$, a weight proportional to the product of the expansion coefficient $\overline{\Lambda}_{u,M}^{-1}(x_j)$ times the inverse of the sine of the angle $\vartheta(S^{M/2} x_j)$ formed by the stable and unstable manifolds at $S^{M/2} x_j$: $\delta_M(x_j) \stackrel{def}{=} \sin \vartheta(S^{M/2} x_j)$. So that the integral of a smooth function $F$ is:

$$\int_{\mathcal{C}} \mu_M(dx) F(x) \stackrel{def}{=} \frac{\sum_j \overline{\Lambda}_{u,M}^{-1} \delta_M^{-1}(x_j) F(x_j)}{\sum_j \overline{\Lambda}_{u,M}^{-1}(x_j) \delta_M^{-1}(x_j)} \quad (4.2)$$

This is interesting because the following theorem holds:

*Theorem: If $(\mathcal{C}, S)$ is a transitive Anosov system the SRB distribution $\mu$ exists and the $\mu$ average of a local function $F$ is:*

$$\int_{\mathcal{C}} \mu(dx) F(x) = \lim_{M \to \infty} \int_{\mathcal{C}} \mu_M(dx) F(x) \qquad (4.3)$$

*and "local" means that $F$ depends "exponentially little" on the symbols with large time label in the symbolic representation of the phase space points (see above; i.e. $|F(x) - F(y)|$ is exponentially small as $k \to \infty$ if the histories of $x, y$ at the times between $-k$ and $k$).*

The above is a trivial reformulation of a deep result of Sinai. It was pointed out in [G1],[G2] and used in [GC1], [GC2]. I will call it *Sinai's theorem*.

The original statement is that the SRB distribution $\mu$ exists and it is a Gibbs state with potential $\log \Lambda_u^{-1}(x)$: see [Bo],[R2],[S] for a discussion of this form of the statement.

The connection between [S] and the above formulation is discussed in [G1] where (4.3) is discussed with $\mu_M$ defined as in (4.2) *without* the factors $\delta_M^{-1}(x_j)$. In spite of the apparently strong modification the extra factor $\delta_M^{-1}(x_j)$. introduces, it is easily seen that (4.3) is valid by examining the proof of (4.3), see [G1], [G2].

The proof is based on the interpretation of (4.2) as a probability distribution on the space of the compatible strings. In this interpretation one immediately recognizes that (4.2) corresponds to a finite volume Gibbs distribution for a suitable short range hamiltonian defined on the space of compatible strings. The extra factor $\delta_M(x_j)$ corresponds to considering the same Gibbs distribution *just with a different boundary condition*: this becomes irrelevant in the limit as $M \to \infty$ because one dimensional Gibbs states with short range interactions do not have phase transitions and therefore are i insensitive to changes in the boundary conditions.

The choice (4.2) as an approximating distribution for $\mu$ is much better than the one without the factors $\delta_M(x_j)^{-1}$ because it leads to simpler formulae and arguments: we shall call (4.2) a *balanced* approximation to the SRB distribution because as we shall see it is reminiscent of a probability distribution verifying the detailed balance (which however is *not* verified in our models, except in 0 forcing, *i.e.* in equilibrium).

In the case in which the system is reversible (as the model in §2), *i.e.* when there is an isometric map $i : \mathcal{C} \to \mathcal{C}$ such that $iS = S^{-1} i$ and $i^2 = 1$, one can add to all the above properties the further properties that $\mathcal{E}$ can be taken *time reversible*. This means that if $E \in \mathcal{E}_M$ then also $iE \in \mathcal{E}_M$, with $i$ being the time reversal operation, and $ix_E = x_{iE}$. Furthermore, recalling (1.2), the following symmetry holds:

$$\sigma_M(x) = -\sigma_M(ix), \qquad \overline{\Lambda}_{u,M}(ix) = \overline{\Lambda}_{s,M}^{-1}(x), \qquad \delta_0(ix) = \delta_0(x), \qquad \delta_M(ix) = \delta_{-M}(x) \quad (4.4)$$

which are identities (see [GC2]) simply implied by the definition of $\sigma$ or by the fact that the $i$ operation changes the stable manifold for $S$ at $x$ into the unstable manifold at $ix$ (still for for $S$).

Furthermore the following relation holds between $\Lambda_u, \Lambda_s, \delta, \sigma$:

$$\Lambda_u(x) \Lambda_s(x) \frac{\delta(Sx)}{\delta(x)} = e^{-\sigma(x)}, \qquad \overline{\Lambda}_{u,M}(x) \overline{\Lambda}_{s,M}(x) \frac{\delta_M(x)}{\delta_{-M}(x)} = e^{-M\sigma_M(x)} = e^{-M\langle \sigma \rangle + p} \qquad (4.5)$$

if one recalls the definitions (1.2),(1.3); the second relation follows from the first: the l.h.s. is in fact the phase space contraction under the map $S^M$ evaluated at the point $S^{-M/2} x$.

For small dimensionality systems the sets $E_q$ can be quite easily and effectively constructed: but the construction, which involves among other things, solving the equations of motion for many data, becomes quickly practically impossible, [FZ].

### §5. Applications of the Chaotic hypothesis. Fluctuation theorem for the KO model.

We study the KO model in (2.3), or the modification considered at the end of §2, and we suppose that the system is kept in a stationary state at a constant energy $U$ under the action of a force field $\underline{g}$ and of the thermostatting mechanism provided by the terms $-\alpha\underline{\gamma}_k$ in the equations of motion.

We study the probability that the fluctuaton variable $p$ is in a small interval $I_q = [q, q+dq]$. We use the notations and the approximation $\mu_\tau$ to $\mu$ described at the end of §4 (see (4.2)) with $F(x) = \sigma_\tau(x)$; the probability that $p \in I_q$ over the probability that $p \in I_{-q}$ is, for large $\tau$:

$$\frac{\sum_{j,p\in I_q} \overline{\Lambda}_{u,\tau}^{-1}(x_j)\delta_\tau^{-1}(x_j)}{\sum_{j,p\in I_{-q}} \overline{\Lambda}_{u,\tau}^{-1}(x_j)\delta_\tau^{-1}(x_j)} \tag{5.1}$$

Since $\mu_\tau$ in (4.2) is only an approximation at fixed $\tau$ *an error is involved in using (5.1)*. It can be shown that this error can be estimated to affect the result only by a factor bounded above and below uniformly in $\tau, p$, [GC1], [G1], [G2]. This is a remark technically based on the proof of the theorem quoted in §4 (which relates the problem to the properties of a one dimensional short range Ising chain, a technical tool that is usually called the "thermodynamic formalism") and it is valid in general for any system verifying the chaotic hypothesis, *i.e.* for any reversible transitive Anosov diffeomorphism, [G2].

It is now possible by using the reversibility to establish a one to one correspondence between the addends in the numerator of (5.1) and the ones in the denominator, (aiming at showing that corresponding addends have a *constant ratio* which will, therefore, be the value of the ratio in (5.1)).

The ratio (5.1) can be written simply as:

$$\frac{\sum_{E_j,p\in I_q} \overline{\Lambda}_{u,\tau}^{-1}(x_j)\delta_\tau^{-1}(x_j)}{\sum_{E_j,p\in I_{-q}} \overline{\Lambda}_{u,\tau}^{-1}(x_j)\delta_\tau^{-1}(x_j)} \equiv \frac{\sum_{E_j,p\in I_q} \overline{\Lambda}_{u,\tau}^{-1}(x_j)\delta_\tau^{-1}(x_j)}{\sum_{E_j,p\in I_q} \overline{\Lambda}_{s,\tau}^{-1}(x_j)\delta_{-\tau}^{-1}(x_j)} \tag{5.2}$$

where $x_j \in E_j$ is a point in $E_j$. In deducing the second relation we make us of the existence of the time reversal symmetry $i$ and of (4.4).

It follows then that the ratios between corresponding terms in the ratio (5.2) is equal to $\overline{\Lambda}_{u,\tau}^{-1}(x)\overline{\Lambda}_{s,\tau}^{-1}(x)\frac{\delta_\tau^{-1}(x_j)}{\delta_{-\tau}^{-1}(x_j)}$. This is, by (4.5), the reciprocal of the total variation of phase space volume over the $\tau$ time steps, if the evolution is regarded as a map on $\mathcal{C}$, between the point $S^{-\tau/2}x$ and $S^{\tau/2}x$: but the reciprocal of the total phase space volume contraction over a time $\tau$ is $e^{q\langle\sigma\rangle+\tau}$. Hence the ratio (5.1) will be $e^{\tau\langle\sigma\rangle+q}$ proving (1.4).

It is important to note that there is one error ignored here. As pointed out in the discussion above the use of $\mu_\tau$ to evaluate the probability is not immediately justified by the theorem of §4 as the function $F$ of which we study the distribution is not a "local" function, because $\sigma_\tau(x)$ in (1.2) is not localized. In fact it depends on the history of $x$ for a time between $-\tau, \tau$ (hence we essentially need the knowledge of the symbols between $-\tau/2$ and $\tau/2$ in the history of $x$ to compute the value of $\sigma_\tau(x)$). This is a delicate point which has been discussed in detail in [G1],[G2]: showing that this interchange of limits problem is really not a problem at all requires going into the details of the proof of the theorem of §4.

One should note that other errors may arise because of the approximate validity of our main chaotic assumption (which states that things go "as if" the system was Anosov): they may depend on $R$, *i.e.* on the number of degrees of freedom, and we do not control them except for the fact that, if present, their relative value should tend to 0 as $R \to \infty$.

The $p$ independence of the limit in (1.4) is therefore a key test of the theory (for a Anoson system; and from the detailed estimates in [G2] one sees that the limit is reached as $\tau \to \infty$ with corrections of order $O(\tau^{-1})$, for $p$ in a fixed bounded interval).

### §6. Reciprocity in a model for a shell motion.

In [CG] it will be shown that the chaotic hypothesis implies also, in a variety of non equilibrium statistical mechanics reversible models, the Onsager reciprocity relations.

It is tempting to apply the above ideas to the models, so far considered, for fluids in states of developed turbulence. There is, however, a basic difficulty: namely the above mentioned derivations deal with "infinitesimal deviations from equilibrium": because they express properties of the second derivatives of the dissipation rate with respect to the thermodynamic forces *evaluated at 0 thermodynamic forces*.

The fluid models considered above (KO and its variation mentioned in §2) do not fulfill the condition of being close to equilibrium: in fact we always imagine $R$ to be very large so that the Kolmogorov Obuchov theory applies and the models may be regarded as describing chaotic motions and as physically reasonable.

There is no known extension of Onsager reciprocity to strongly forced systems, [DGM]: therefore we shall not insist in studying the above models. But a possible application to fluid motions can be found by considering a related model.

We consider a fluid whose velocity field contains only components with momenta in the range $\mathcal{M}_n$ defined by $2^{n-1} k_0 \le |\underline{k}| \le 2^n k_0$ containing $N$ pairs $\underline{k}, -\underline{k}$: this is often called a *momentum shell*. The equations of motion will be:

$$\dot{\underline{\gamma}}_{\underline{k}} = -i \sum_{\underline{k}_1 + \underline{k}_2 = \underline{k}} (\underline{\gamma}_{\underline{k}_1} \cdot \underline{k}_2) \Pi_{\underline{k}} \gamma_{\underline{k}_1} + \underline{f}_{\underline{k}} - \alpha \underline{\gamma}_{\underline{k}} \qquad 2^{n-1} k_0 \le |\underline{k}|, |\underline{k}_1|, |\underline{k}_2|) \le 2^n k_0 \quad (6.1)$$

and $\alpha$ is fixed so that the total kinetic energy is $2NkT$, hence the relations between kinetic energy, the "friction" $\alpha$, and the phase space contraction exponent $\sigma$ are:

$$\frac{1}{2} \sum_{\underline{k}} |\underline{\gamma}_{\underline{k}}|^2 = 2NkT, \qquad \alpha = \frac{\sum_{\underline{k}} \underline{f}_{\underline{k}} \cdot \underline{\gamma}_{\underline{k}}}{2NkT}, \qquad \sigma = \frac{\sum_{\underline{k}} \underline{f}_{\underline{k}} \cdot \underline{\gamma}_{\underline{k}}}{kT/2} t_0 \qquad (6.2)$$

if $t_0$ is the average time interval between timing events. As already mentioned after (2.5), *for simplicity of exposition*, we neglect that there is a difference between the actual time $t(x)$ elapsing between the timing event $x \in \mathcal{C}$ and the successive $Sx$ and that during such intervals $\sigma(V_t x)$ is not strictly constant. In fact $\sigma$ should be really written as an integral over the continuous time trajectory $t \to V_t x$ described in the time $t(x)$ while the point $x$ evolves between $x$ and $Sx$.

One can think that this is a crude model for the motion of the velocity field components with momenta of scale $n$, if the range $\mathcal{M}_n$ is in the inertial range of a strongly turbulent flow.

In this model we can ask the question of the reciprocity of the fluctuation responses with respect to the variations of the forces $\underline{f}_{\underline{k}}$.

Denoting $\beta$ a pair $\underline{k}, h$ with $h = 1, 2, 3$, and setting $f_\beta \equiv f_{\underline{k}, h}$ and $\partial_\beta = \frac{\partial}{\partial f_{\underline{k}, \beta}}$, we want to show, under an extra assumption (see the two paragraphs following (6.14) below), that the chaoticity hypothesis implies:

$$\partial_{\beta'}\langle\partial_{\beta}\sigma\rangle_{\mu}\Big|_{\underline{f}=\underline{0}} \stackrel{def}{=} L_{\beta,\beta'} = L_{\beta',\beta} \stackrel{def}{=} \partial_{\beta}\langle\partial'_{\beta}\sigma\rangle_{\mu}\Big|_{\underline{f}=\underline{0}} \tag{6.3}$$

where $\mu$ denotes the SRB distribution.

In spite of its appearance (6.3) requires some discussion on its meaning.

The function $t_0^{-1}\sigma(x)$ is, by definition, the phase space entropy production rate. it is naturally defined on the phase space $\mathcal{C}$. However if $\langle\sigma\rangle_+ > 0$ (as it is the case for $\underline{f} \neq \underline{0}$) the motion is dissipative and the attractor for the motion has zero volume, see [GC2]. In fact we expect, believing the pairing rule discovered in [ECM1], see also [GC2], that the fractal dimension of $A$ is macroscopically lower than that of $\mathcal{C}$.

Therefore the "relevant" values of the function $\sigma(x)$ are those for which $x \in A$. But $A$ as well as the SRB distribution $\mu$ are $\underline{f}$-dependent. Hence if we want to discuss the $\underline{f}$ dependence of $\sigma$ we must think that $\sigma$ is defined on a surface that generates the attractor, *e.g.* the unstable manifold of a fixed point $O$ (or periodic motion) $W^u(O)$, and *which depends on $\underline{f}$*.

The consequence is that $\partial_{\beta}(\sigma(x))$ *cannot* be identified with the partial derivative of (6.2) (*i.e.* $\frac{2t_0}{kT}\gamma_{\beta}$) but one has *also* a contribution $\frac{\partial\sigma(x)}{\partial x} \cdot \frac{\partial x}{\partial f_{\beta}}$ as the phase space point $x$ has to be fixed on the attractor and it *changes* therefore with $\underline{f}$.

A simple way to see this from a different point of view is to think of the attractor as represented via the symbolic strings $\underline{j}$ associated with a Markov partition: $x = x(\underline{j})$, as discussed in §4.

The SRB distribution becomes a probability distribution over the family of compatible strings $\underline{j}$ and, in fact, it is a Gibbs state corresponding to a potential which has short range (*i.e.* by a potential with energy per site given by the function $\log\overline{\Lambda}_u^{-1}(x(\underline{j}))$).

In this language the *dynamics becomes $\underline{f}$-independent* and $\sigma$ becomes a function of the string $\sigma = \sigma(x(\underline{j}))$. Therefore $\sigma$ depends on $\underline{f}$ for *two* reasons:

(1) $\sigma$ depends "explicitly" on $\underline{f}$, see (6.2).
(2) $x(\underline{j})$ depends on $\underline{f}$ as well, because the correspondence between strings $\underline{j}$ and points $x(\underline{j})$ depends on the dynamics.

Hence we shall use the more appropriate notation: $\sigma(x) = s(\underline{f}, x(\underline{j}))$ and:

$$\frac{\partial\sigma}{\partial f_{\beta}} = \frac{\partial s}{\partial f_{\beta}}(\underline{f}, x(\underline{j})) + \frac{\partial s}{\partial x}(\underline{f}, x(\underline{j})) \cdot \frac{\partial x(\underline{j})}{\partial f_{\beta}} \tag{6.4}$$

Of course it is $s(\underline{0}, x(\underline{j})) = 0$ so that we get:

$$\frac{\partial\sigma}{\partial f_{\beta}}\Big|_{\underline{f}=\underline{0}} \equiv \frac{\partial s}{\partial f_{\beta}}\Big|_{\underline{f}=\underline{0}} \tag{6.5}$$

The proper interpretation of (6.3) is obtained by defining $\partial_{\beta}$ as in (6.4), which makes a difference even when we may only be interested in evaluating $L_{\beta\beta'}$ at $\underline{f} = \underline{0}$. In fact (6.3) involves the *second* derivative of $\sigma$, for which a relation like (6.5) *does not necessarily hold*.

We shall see below that (6.3) holds with the latter interpretation of $\partial_{\beta}\sigma$. Hence we argue that the correct definition of the *current* $j_{\beta}$ when $\underline{f} \neq \underline{0}$ seems to be:

$$j_{\beta} = \frac{1}{t_0}\langle\partial_{\beta}\sigma\rangle \tag{6.6}$$

where $\partial_{\beta}\sigma$ is defined as in (6.4).

One could also object that since the code $\underline{j} \to x(\underline{j})$ depends on $\underline{f}$ it might happen that the space of the compatible sequences itself changes with $\underline{f}$: *i.e.* that the compatibility matrix depends on $\underline{f}$ introducing a further dependence of $\sigma$ on $x$ and in fact a dependence on $\underline{f}$ on the symbolic dynamics itself. Although this might indeed happen at "large" (perhaps moderately large) values of $\underline{f}$ it is a consequence of Anosov structural stability theorem, see [AA], that the compatibility matrix $C$ for the compatible strings $\underline{j}$ *does not* change for small enough variations of the dynamics, hence of $\underline{f}$. In fact establishing the constancy of the stability matrix is the first step in the proof of the structural stability (and essentially the reason for the stability itself).

The above definition (6.6) of the notion of current associated with a thermodynamic force seems to be new and it should be carefully understood and tested in the nonlinear regime. Here we assume *by definition* that the current associated with the thermodynamic force component $f_\beta$ by (6.6) with the derivative of $\sigma$ defined by (6.4).

The proof of (6.3) now follows a natural scheme, see also [CG]. We set, in an effort to simplify notations in the coming formulae, $l_{u,\tau}, l_{s,\tau}$ as:

$$l_{u,\tau}(x) = \log \overline{\Lambda}_{u,\tau}^{-1}(x)\delta_\tau^{-1}(x), \qquad l_{s,\tau}(x) = \log \overline{\Lambda}_{s,\tau}(x)\delta_{-\tau}^{-1}(x) \tag{6.7}$$

so that by (4.4):

$$l_{u,\tau}(x) - l_{s,\tau}(x) = \tau \sigma_\tau(x), \qquad l_{u,\tau}(ix) = l_{s,\tau}(x) + \tau \sigma_\tau(x) \tag{6.8}$$

and we see that $j_\beta = \frac{1}{t_0}\langle \partial_\beta \sigma \rangle$ (see (6.4)) is the limit of the r.h.s. of (4.2) with $F(x) = \partial_\beta \sigma(x)$ and $M = \tau$ as $\tau \to \infty$. We can also say that $\langle \partial_\beta \sigma \rangle$ is the limit of (4.2) with $F(x) = \partial_\beta \sigma_\tau(x)$, replacing $\sigma$ by its average, see (1.2), between $-\frac{1}{2}\tau$ and $\frac{1}{2}\tau$. This is easily justified if one recalls the symbolic dynamics interpretation of $\mu_\tau$ (discussed after the theorem in §4) in terms of a 1–dimensional Gibbs distribution for a short range Ising model (see the second comment after (4.3)). I will not discuss this (minor) technical point further. Hence:

$$\begin{aligned}\lim_{\tau \to \infty} t_0 \partial_{\beta'} j_\beta &= \frac{\sum_j \overline{\Lambda}_{u,\tau}^{-1}(x_j)\delta_\tau^{-1}(x_j)\Big(\partial_{\beta'\beta}\sigma_\tau(x_j) + \partial_{\beta'}l_{u,\tau}(x_j)\partial_\beta \sigma_\tau(x_j)\Big)}{\sum_j \overline{\Lambda}_{u,\tau}^{-1}(x_j)\delta_\tau^{-1}(x_j)} - \\ &\quad - \frac{\sum_j \overline{\Lambda}_{u,\tau}^{-1}(x_j)\delta_\tau^{-1}(x_j)\partial_{\beta'}l_{u\tau}(x_j)}{\sum_j \overline{\Lambda}_{u,\tau}^{-1}(x_j)\delta_\tau^{-1}(x_j)} \cdot \frac{\sum_j \overline{\Lambda}_{u,\tau}^{-1}(x_j)\delta_\tau^{-1}(x_j)\partial_\beta \sigma_\tau(x_j)}{\sum_j \overline{\Lambda}_{u,\tau}^{-1}(x_j)\delta_\tau^{-1}(x_j)} = \\ &= \lim_{\tau \to \infty}\Big(\langle \partial_{\beta'\beta}\sigma_\tau\rangle + \big(\langle \partial_{\beta'}l_{u,\tau}(x_j)\,\partial_\beta \sigma_\tau\rangle - \langle \partial_{\beta'}l_{u,\tau}\rangle\langle \partial_\beta \sigma_\tau\rangle\big)\Big)\end{aligned} \tag{6.9}$$

By using the time reversal invariance we see that:

$$\langle \partial_{\beta'}l_{u,\tau}\partial_\beta \sigma_\tau\rangle = \frac{\sum \overline{\Lambda}_{u,\tau}^{-1}\delta_\tau^{-1}(x_j)\partial_{\beta'}l_{u,\tau}\,\partial_\beta \sigma_\tau}{Z} = \tag{6.10}$$

$$= \frac{\sum_j \big(\overline{\Lambda}_{u,\tau}^{-1}(x_j)\delta_\tau^{-1}(x_j)\partial_{\beta'}l_{u,\tau}(x_j)\,\partial_\beta \sigma_\tau(x_j) + \overline{\Lambda}_{u,\tau}^{-1}(ix_j)\delta_\tau^{-1}(ix_j)\partial_{\beta'}l_{u,\tau}(ix_j)\,\partial_\beta \sigma_\tau(ix_j)\big)}{2Z}$$

where $Z$ denotes the "partition sum", *i.e.* the sum in the denominator of (6.9), and the averages are with respect to the distribution $\mu_{\tau/2,\tau}$. Recalling (6.7) this becomes:

$$\frac{\sum_j \big(\overline{\Lambda}_{u,\tau}^{-1}(x_j)\delta_\tau^{-1}(x_j)\partial_{\beta'}l_{u,\tau}(x_j) - \overline{\Lambda}_{s,\tau}(x_j)\delta_{-\tau}^{-1}(x_j)\partial_{\beta'}l_{s,\tau}(x_j)\big)\,\partial_\beta \sigma_\tau(x_j)}{2Z} \tag{6.11}$$

Equation (6.8) permits us to reconstruct $\tau \partial_{\beta'}\sigma_\tau$ ou of the two addends in (6.11). And the derivatives at $\underline{f} = \underline{0}$ can be computed immediately by noting that *in such case*, by (4.5), (6.7),(6.8):

$$\Big(\langle \partial_{\beta'}l_{u,\tau}\, \partial_\beta \sigma_\tau\rangle - \langle \partial_{\beta'}l_{u,\tau}\rangle \langle \partial_\beta \sigma_\tau\rangle\Big)\Big|_{\underline{f}=\underline{0}} = \frac{\tau}{2}\Big(\langle \partial_{\beta'}\sigma_\tau\, \partial_\beta \sigma_\tau\rangle - \langle \partial_{\beta'}\sigma_\tau\rangle \langle \partial_\beta \sigma_\tau\rangle\Big)\Big|_{\underline{f}=\underline{0}} \qquad (6.12)$$

The quantity $\langle \partial_{\beta\beta'}\sigma_\tau\rangle$ will converge to a limit $\langle \partial_{b\beta'}\sigma\rangle$ because the SRB distribution is stationary, neglecting an exchange of limit (see comments below); then by using (1.2):

$$\begin{aligned}
\partial_{\beta'}j_\beta\Big|_{\underline{f}=\underline{0}} &= \lim_{\tau\to\infty}\Big(\frac{1}{t_0}\langle \partial_{\beta\beta'}\sigma\rangle\Big|_{\underline{f}=\underline{0}} + \\
&+ \frac{1}{2\tau t_0}\sum_{m=-\tau/2}^{\tau/2-1}\sum_{n=-\tau/2}^{\tau/2-1}\big(\langle \partial_{\beta'}\sigma(S^m\cdot)\partial_\beta \sigma(S^n\cdot)\rangle - \langle \partial_{\beta'}\sigma(S^m\cdot)\rangle\langle \partial_\beta\sigma(S^n\cdot)\rangle\big)\Big|_{\underline{f}=\underline{0}} = \\
&= \frac{1}{t_0}\langle \partial_{\beta\beta'}\sigma\rangle\Big|_{\underline{f}=\underline{0}} + \frac{1}{2t_0}\sum_{m=-\infty}^{\infty}\big(\langle \partial_{\beta'}\sigma(S^m\cdot)\partial_\beta\sigma(\cdot)\rangle - \langle \partial_{\beta'}\sigma(\cdot)\rangle\langle \partial_\beta\sigma(\cdot)\rangle\big)\Big|_{\underline{f}=\underline{0}}
\end{aligned} \qquad (6.13)$$

where the averages in the r.h.s. are with respect to $\mu = \lim \mu_\tau$. This, again apart from a problem of interchange of limits (see comments below) becomes:

$$\begin{aligned}
\partial_{\beta'}j_\beta\Big|_{\underline{f}=\underline{0}} &= \frac{1}{2t_0}\sum_{m=-\infty}^{\infty}\big(\langle \partial_k \beta' \sigma(S^m\cdot)\partial_\beta\sigma(\cdot)\rangle - \langle \partial_{\beta'}\sigma(\cdot)\rangle\langle \partial_\beta\sigma(\cdot)\rangle\big)\Big|_{\underline{f}=\underline{0}} + \\
&+ \frac{1}{t_0}\langle \partial_{\beta\beta'}\sigma\rangle = \frac{1}{2t_0}\sum_{m=-\infty}^{\infty}\langle \partial_{\beta'}\sigma(S^m\cdot)\partial_\beta\sigma(\cdot)\rangle
\end{aligned} \qquad (6.14)$$

where the averages are with respect to the SRB distribution (*i.e.* with respect to the limit of $\mu_\tau$) and the missing terms vanish because of time reversal symmetry (for instance $\frac{1}{t_0}\langle \partial_{\beta\beta'}\sigma\rangle$ is seen, considering (6.2) and (6.4), to coincide with the expectations $\langle \partial_\beta\gamma_{\beta'} + \partial_{\beta'}\gamma_\beta\rangle$, linear in $\gamma$ and therefore vanishing by the time reversal symmetry).

The problems of interchange of limits are easily solved: under our assumption that the system is a transitive Anosov system the correlations of smooth observables decay exponentially (because they become local observables in the symbolic dynamics interpretation of the evolution, provided by the Markov partitions), not only for $\mu$ but also for $\mu_\tau$ (in the natural sense in which this may be interpreted in a finite $\tau$ case; *e.g.* by regarding the interval $[-\frac{\tau}{2}, \frac{\tau}{2}]$ as a circle), and uniformly in $\tau$.

Note that here there is one hidden assumption: namely the "local observable" to which we want to apply the above argument is $\partial_\beta\sigma$ which, by (6.8) contains besides the really local part given by the first term in the r.h.s. a second part containing $\partial_\beta x(\underline{j})$ (and $\partial_\beta x(\vartheta^k\,\underline{j})$ with $\vartheta$ being the shift map on the symbolic sequences): therefore we must assume that also $\partial_\beta x(\underline{j})$ is "local" in the sense of the theorem in §4, which seems a reasonable assumption. One can *conjecture* that it holds in general for any transitive Anosov system (or axiom A system) depending smoothly on a parameter $f_\beta$.

This also shows that (6.14) is indeed correct and, therefore, recalling that $L_{\beta\beta'} = t_0 \partial_{\beta'}j_\beta$ we get (6.3) as well as the important property that the matrix $L_{\beta\beta'}$ is *positive definite* (see the last of (6.14)). And one can check (by (6.2) and (6.4)), that the relation $L_{\beta\beta'} = L_{\beta'\beta}$ is, in fact:

$$\partial_{\beta'}\langle \gamma_\beta\rangle\Big|_{\underline{f}=\underline{0}} = \partial_\beta \langle \gamma_{\beta'}\rangle\Big|_{\underline{f}=\underline{0}} \qquad (6.15)$$

which is easier to interpret.

*Remarks:*

(1) Note that (6.13) seems to have one "extra term", $\frac{1}{t_0}\langle \partial_{\beta\beta'}\sigma\rangle$, with respect to what one usually expects from linear response theory: this is due to the fractality of the attractor and it arises because od the "extra term" in (6.4).

(2) If one thinks that the above is a reasonable model for the evolution of the velocity components realtive to a shell in the inertial range, then above relation should be subjected to an experimental test. In any event the relations are very likely testable in numerical experiments on the solutions of the equations (6.1).

(3) One can remark that if $\gamma_\beta \equiv \gamma_{\underline{k},h}$ for $\beta = (\underline{k},h)$ the above reciprocity relations can be written as the symmetry of the matrix $\partial_\beta \langle \gamma_{\beta'}\rangle_\mu\big|_{\underline{\varphi}=\underline{0}}$ *only* if we consider transformations of the system which vary $\underline{f}$ but keep the total energy $U = NkT$ constant.

(4) One certainly wants to consider also transformations in which $U$ changes according to some *equation of state* $U = g(\underline{f})$. The function $g$ depends on the physical situations, *i.e.* on the actual mechanism of dissipation that generates the model (6.3): however if the model is regarded only as a mathematical fiction, then of course the equation of state is arbitrary. In this case, given $g$, the above derivation of the reciprocity relations still holds up to some natural modifications, but the symmetry $\partial_{\beta'}\langle\partial_\beta\sigma\rangle$ does not directly mean the symmetry (6.15). One can, however, check that it implies at $\underline{f} = \underline{0}$:

$$\partial_{\beta'}(\langle\gamma_\beta - \frac{\sigma}{T}\partial_\beta U\rangle) = \partial_\beta(\langle\gamma_{\beta'} - \frac{\sigma}{T}\partial_\beta U\rangle) \tag{6.16}$$

which still implies (6.15) (again by the time reversal symmetry). This is in agreement with the philosophy exposed in §3.

*Acknowledgements:* I am indebted to J.L. Lebowitz and G.L. Eyink for very helpful comments and to E.G.D Cohen for constant encouragement and many very helpful suggestions and comments. This work is part of the research program of the European Network on : "Stability and Universality in Classical Mechanics", # ERBCHRXCT940460.